\documentclass[prb,amsmath,amssymb]{revtex4}

\usepackage{graphicx}

\def\la{\langle}
\def\ra{\rangle}

\begin{document}



\title{Quantum Noise, Detailed Balance and Kubo Formula \\ in Nonequilibrium Quantum Systems}

\author{{U. Gavish}, \underline{Y. Imry}, Y. Levinson}
\affiliation{ Weizmann Institute of Science \\ 76100 Rehovot,
Israel}

\begin{abstract}
Current quantum noise can be pictured as a sum over transitions
through which the electronic system exchanges energy with its
environment. We formulate this picture and use it to show which
type of current correlators are measurable, and in what
measurement the zero point fluctuations will play a role (the
answer to the latter is as expected: only if the detector excites
the system.) Using the above picture, we calculate and give
physical interpretation of the finite-frequency finite-temperature
current noise in a noninteracting Landauer-type system,  where the
chemical potentials of terminals 1 and 2 are $\mu+eV/2$ and
$\mu-eV/2$ respectively, and derive a detailed-balance condition
for this nonequilibrium system. Finally, we derive a generalized
form of  the Kubo formula for a wide class of interacting
nonequilibrium systems, relating the \textit{differential}
conductivity to the current noise.
\end{abstract}

\maketitle Contribution to the Proceedings of the 2001 Recontres
de Moriond: Electronic Correlations: from Meso- to Nanophysics, T.
Martin, G. Montambaux and J. Tran Thanh Van eds. EDPScience 2001.

\section{Introduction}

A  general expression for the current correlators is derived
below: it is a sum over transitions between pairs of energy
levels. This expression is very similar to the one obtained by Van
Hove \cite{VH} in the framework of neutron scattering theory.
Using this sum over transitions we will identify the positive and
negative frequency part of the correlator as having distinct,
well-defined, separately-measurable, physical interpretations,
namely: the emission and absorption spectra. This picture will
enable us to analyze in what type of measurement the zero-point
fluctuations will have an effect on the measured spectrum. It will
also enable us to derive a detailed balance relation out of
equilibrium for a Landauer-type system which consists of a point
scatterer connected through two one-dimensional single-channel
conducting arms, to two terminals whose chemical potentials
differ. This condition keeps the system in a stationary
nonequilibrium state. Finally, a Kubo formula for the differential
conductivity is derived for a wide class of systems out of
equilbrium, and is verified analytically for the above
Landauer-type case.

\section {General Formulation for the Quantum noise.}
\label{general}

Consider the current correlator for a stationary system (i.e. with
no dependence on $t'$),
\begin{eqnarray}
\label{Cq} C(t)\equiv\la \hat J(t')\hat J(t' + t)\ra\equiv\sum_{i}
P_{i}\la i|\hat J(0)\hat J(t)|i\ra ,
\end{eqnarray}
of a quantum system ("antenna"), characterized by a density matrix
which is diagonal in the eigenstate basis. $|i\ra$ are the
eigenstates of the antenna with energies $E_{i}$ and populations
$P_{i}$. $\hat J$ is a space average \cite{jq} of the
time-dependent current operator in the Heisenberg representation,
$\hat j(x,t) =\exp(i\hat H t/\hbar)\hat j \exp(-i\hat H t/\hbar)$,
where $\hat H$ is the Hamiltonian of the antenna, taken to be time
independent, but otherwise very general, {\em including
interactions}.

If the (stationary) system were describable classically, then one
would have $C(t) = C(-t)$
 and it would be clear what the time-dependent current fluctuations were:
 the average-over-realizations of the product of the
values of the current at different times. However, in the quantum
case the operators $\hat j(t)$ for different times do not commute.
Therefore the quantum correlator can not, in general, be given
this simple interpretation.
 Actually, in the quantum case $C(t)$ is in general \textit{complex}
 and not symmetric, but just satisfies: $ C(t)=C(-t)^*,$
 which means that \textit{it is not a directly measurable quantity} in the sense that
the product of the (real) current values that appear, say, on some
ampermeter screen at
 different times does not give $C(t)$.
Its Fourier transform (to which we shall also refer, for brevity,
as the correlator):
\begin{equation} S(\omega)\equiv \int_{-\infty}^{\infty} e^{i\omega t}\langle |\hat
J(0) \hat J(t)|\rangle dt , \label{defS}
\end{equation}
 is also nonsymmetric,
$S(\omega)\neq S(-\omega),$ although it is obviously real (see
eq.(\ref{Sq})). This can be seen\cite{VH} by inserting the
identity operator $\sum_f |f\rangle \langle f|$ in eq.(\ref{defS})
 and performing the time-integration, which yields:
\begin{eqnarray}
\label{Sq}
S(\omega)=2\pi\hbar\sum_{if} P_{i}|\la f|\hat J|i\ra|^2
\delta(E_{i}-E_{f}-\hbar\omega).
\end{eqnarray}
This expression is not symmetric with respect to $\omega$ since
$P_{i}$ usually decreases with $E_{i}$.

$S(\omega)$ has the following important physical significance. If the system is
coupled, through a small term which is linear in $\hat J(t)$, to a second system
(see examples below),
 then, by the Fermi golden rule, $S(\omega)$ is proportional to the transition rate between the
initial state $|i\rangle$ and the final state $|f\rangle$, for
which $E_i-E_f=\hbar\omega$. Therefore, for $\omega > 0$ it is
proportional to the emission rate and for $\omega < 0$ it is
proportional to the  absorption rate \cite{power}.

For example, if the second system is the free EM field, then
 $S(\omega>0)$  is proportional to the energy emission rate \cite{power,Baym} into the
vacuum state of this field (i.e. the state of the EM field with occupation number $N_{\omega}= 0$),
 and for $\omega < 0$ it is proportional to the energy absorption rate from an EM field
with a given photon, i.e., with $N_{|\omega|} = 1$. Another
example is provided by a system which is coupled to a measuring
device \cite{LES1,T0,LI,Leo} (e.g., a resonant circuit). Then
$S(\omega)$ is proportional to the energy transfer rate between
the system and the measuring-device: The terms with $E_{i}>E_{f}$
describe transitions in which an energy of $\hbar\omega = E_{i} -
E_{f}
>0$ is transferred from the system to the measuring device, while terms
with $ E_{f}>E_{i}$ describe
transitions in which an energy of $-\hbar\omega = E_{f} - E_{i}
>0$ is transferred from the measuring device to the system.
When $\omega>0$, only the first type of terms will
remain and $S(\omega)$ will be the emission spectrum while
$S(-\omega)$ will be the absorption spectrum.
Thus, the two branches of $S(\omega)$ yield two physically
interesting and separately-measurable quantities.

In the case when the antenna is in equilibrium at a temperature
$T$ and time-reversal symmetry holds, one finds the detailed
balance relation \cite{VH}
\begin{equation}
S(\omega) = S(-\omega) e^{- \hbar\omega /k_{B} T}.
\label{db}
\end{equation}

This detailed balance relation ensures that the system remains in
equilibrium, by taking care that the asymmetry $S(\omega)\not=
S(-\omega)$, i.e., the difference between the upward transitions
(absorption) and the downward ones (emission) is compensated by
the difference between the higher and lower thermal occupations.
In section 3 the above relation is generalized for a particular
nonequilibrium system, where it serves to keep the latter in its
nonequilibrium, though stationary, state. From eq.(\ref{db}) one
sees that only for low frequencies $\hbar|\omega|\ll k_{B}T$, will
the classical symmetry, $S(\omega) = S(-\omega)$, hold. In the
time domain this means that the classical symmetry, $C(t) =
C(-t)$, becomes valid only for late times $|t|\gg \hbar/k_{B}T$.

The customary way to treat the quantum case is \cite{LES1,LL} to
consider  the symmetrized correlator $C_{S}(t'-t)\equiv(1/2)\la
\hat j(t)\hat j(t')+\hat j(t')\hat j(t)\ra,$ which is real and
symmetric like the classical one. However, for a wide class of
noise detection schemes and in particular for a detector in its
ground-state $C_{S}(t)$ is {\em not},
 the measured correlator,
since it contains the zero-point fluctuations.
For example, if the antenna is in equilibrium at a temperature $T$,
it follows from the fluctuation-dissipation theorem
\cite{LL} that for $\omega >0$ one has
$S_{S}(\omega)\sim [N_{T}(\omega)+(1/2)]\omega,$
where $S_{S}(\omega)$ is the Fourier transform of $C_{S}(t)$
and $N_{T}(\omega) = [exp( \omega /k_{B} T) -1]^{-1}$ is the Planck
function.
This means that $S_{S}(\omega)\neq 0$ even when $T=0$, and the antenna is
in its ground state. Since being in the ground state the
antenna can not radiate energy, $S_{S}(\omega)$ can not be
considered as the correlator measured by detecting the radiation \cite{Gard}.

To conclude, the measured quantity will generally not contain the
emission and absorption in a symmetric combination. This was shown
for particular situations of quantum noise measurement in
electronic transport \cite{LES1,Leo,T0,Gard} as well as in quantum
optics \cite{Lax,Gard,Glauber}. The more physical correlator (and
its power spectrum) is the one without symmetrization. Its
transform is given by a sum over transitions - downward ones in
the case of positive frequency, and upward ones in the case of
negative frequency. Each of these two branches has its own
distinct physical significance, the emission and absorption
spectrum, that may in principle be detected separately.

\section{Shot Noise}

We now consider current fluctuations
\cite{KhlusLesovikButtiker,Yang} for the  Landauer model. A
point-like elastic scatterer is connected through two ideal
single-channel conducting ballistic arms to two Fermion
reservoirs, 1, and 2, with chemical potentials $\mu+eV/2$ and
$\mu-eV/2$ respectively. $V$ is the voltage, and it is assumed
that $eV,\hbar\omega,k_BT\ll\mu$. We consider non-interacting
electrons and ignore spin. The single-particle scattering-states
with energy $\epsilon_n=\hbar^2 k^2/2m,$ which corresponds to a
wave that is incoming on arm $\alpha=1,2$, partially reflected
back into it and partially transmitted into the other arms, is:
$\varphi_n(x_\gamma)= L^{-1/2}
[\delta_{\alpha\gamma}e^{-ikx_{\gamma}}+s_{\gamma\alpha}(k)e^{ikx_\gamma}]$.
Here $n\equiv (\alpha,k)$ with $k>0$;  $\alpha, \gamma =1,2$, $L$
is a normalization length, $m$ the electron mass and $x_{\gamma}$
the distance of a point on arm $\gamma$ from the scatterer. $
s_{\alpha \gamma}$ is the element of the unitary scattering
matrix, and it is assumed to be energy independent unless
otherwise stated.
  To specify that a state $\varphi_{n}$
comes from terminal $\alpha,$ we shall write $n \in \alpha.$
 The current operator on arm  $\alpha$ is $$\hat{j}(x_\alpha)=
 -(ie\hbar/2m) \sum_{nn'} \hat{a}_{n}^{\dag} \hat{a}_{n'} \varphi_{n}
 ^\ast \nabla_{\beta}\varphi_{n'}+h.c.,$$ where
$\hat{a}_n$ and  $\hat{a}_n^{\dag}$ are the annihilation
and creation operators of the $\varphi$'s.

We assume that the {\it measured} current is the average
\begin{equation}
\hat{J}(t)\equiv \frac{1}{L_0}\int_{L_0}dx_{2}\hat{j} (x_2)
\label{J}
\end{equation}
over a segment  $L_0$ far away from the scatterer which satisfies:
$L_0 k_F\gg 1$ and $\omega L_0m/(\hbar k_F) \ll 1,$ where $k_F
\equiv \sqrt{2m\mu}$, and $\omega$ is the frequency of the
measured noise which is assumed to satisfy $\omega\ll\mu$. These
conditions ensure that the correlators are independent of the
length and position of the segment $L_0,$; i.e, it has no spatial
dependence, which is not addressed in experiments.

To describe the current noise consider the correlator, given by eq. (\ref{Sq})
We emphasize again that at
least for some types of noise detection, it is eq. (\ref{defS})
and {\em not} its symmetrized version, which gives the measured
noise if the detector is cold enough, i.e., when excitation of the
system by the detector is unlikely so that the absorption is
negligible. The states $|i\rangle$  are given according to the
Landauer picture by the eigenstates of the system (i.e., Slater
determinants) labelled by specifying a set $\{n_i\}$ of the
occupied single-particle states (the scattering states $\varphi_n$
emanating from the two reservoirs):

\begin{eqnarray}
|i \rangle \equiv |\{n_i\} \rangle =\prod_{n_i}  \hat a_{n_i}^{\dag} |vacuum \rangle,
\label{defi}
\end{eqnarray}

and the corresponding probabilities (for a more general derivation
see Ref.~\onlinecite{Hersh}) are:
\begin{eqnarray}
P_i\equiv \frac{1}{Z_1}exp[-\beta\sum_{n\in
1}(\epsilon_{n_i}-(\mu+eV/2))n_i] \times \frac{1}{Z_2}
exp[-\beta\sum_{n\in 2}(\epsilon_{n_i}-(\mu-eV/2))n_i],
\label{defPi}
\end{eqnarray}
where $\beta\equiv (k_BT)^{-1}$ and
 $$Z_\alpha\equiv \sum_{\{n_i\}}exp[-\beta\sum_{n\in \alpha}n_i(\epsilon_{n_i}-(\mu- eV(-1)^\alpha/2))]  $$
 $\alpha=1,2$.
The probabilities $P_i$ correspond to a situation in which the occupations in the
gas in the one-dimensional system are determined by grand-canonical probabilities that depend on
the chemical potential and the temperature
 of the terminals that supply the electrons to the system. From eqs.(\ref{defPi}) and (\ref{defi})
 it follows that at zero temperature $P_i=0$ for any $i$ except for one state, which we
name a \textit{cold transport state} which is given by:
\begin{eqnarray}
|cold\ \ transport \rangle \equiv &\prod&  \hat a_n^{\dag} |vacuum \rangle,
\nonumber\\ &_{n \in 1 ;\epsilon_n \leq \mu+ eV/2}& \nonumber \\ &_{n \in
2 ;\ \epsilon_n \leq \mu-eV/2}&
\label{deftransportgas}
\end{eqnarray}
In this cold transport state all the $\varphi_n$'s are occupied up
to an energy $ \mu-eV/2$ if $n\in 2$ and up to $\mu +eV/2$
($eV\ll\mu$) if $n\in 1$ (see fig. 1). Therefore, this state,
although it is the stationary state with the lowest energy among
those that are made possible by the two terminals, is not a ground
state, and since it carries current, it is not even an equilibrium
state.

In the sum in eq.(\ref{Sq}), the non-diagonal matrix element
$\langle i| \hat J(0) |f \rangle$ is nonzero only if $|f \rangle$
differs from $|i \rangle$ by moving one particle from an occupied
state, $\varphi_n,$ to a  previously unoccupied state,
$\varphi_{n'},$ i.e., $|f \rangle$ is of the form $
\hat{a}_{n'}^{\dag} \hat{a}_n |i \rangle$ (up to a fermionic
factor of $\pm 1$, that will play no role below.) The diagonal
elements $\langle i| \hat J(0) |i \rangle$ appear in a term $\sim
\delta (\hbar\omega)$. In what follows we consider only $\omega
\not= 0$ and therefore neglect this term. In experiments the
integration in eq.(\ref{defS}) is limited by the sampling time of
the experiment, $T_s$, and as a result  $\delta (\hbar\omega)$ is
smoothed into a peak with a width of $\simeq \hbar/T_s$ which
means that the condition $\omega \not=0$ actually means $\omega
T_s\gg 1$. We therefore have:

\begin{eqnarray}
 S(\omega)=2\pi
\sum_i P_i\sum_{nn' }|J_{nn'}^i|^2 \delta  (\epsilon_{n}-\epsilon_{n'} -
\omega). \label{Svanhove-auto}
\end{eqnarray}
where $J_{nn'}^i \equiv \langle i| \hat J(0) \hat{a}_{n'}^{\dag}
\hat{a}_n |i \rangle$, and where now the summation over $n$ and
$n'$ is over all {\it single-particle} states $\varphi _n$ and
$\varphi_{n'}$ (with \textit{single-particle} energies
$\epsilon_{n}$ and $\epsilon_{n'}$)
 which are occupied and unoccupied, respectively, in $|i\rangle$.
Now we divide the summation in eq. (\ref{Svanhove-auto}) into four
partial sums, according to the four possible types of transitions:
two auto-terminal ones and two cross-terminal ones (shown in fig.
1 for $\omega<0$ in the cold transport state):

\begin{eqnarray}
 S(\omega)=\sum_{\alpha,\gamma=1,2}S_{\alpha\rightarrow \gamma}(\omega),
\label{partialsums}
 \end{eqnarray}
 where $S_{\alpha\rightarrow \gamma}(\omega)$ contains only the transitions from
 states with $n\in\alpha$ to states $n\in \gamma$. Explicitly:
\begin{eqnarray}
 S_{\alpha\rightarrow \gamma}(\omega)=2\pi \sum_i P_i\sum_{ \ n\in \alpha,\ \ n'\in\gamma }|J_{ nn'}^i|^2 \delta  (\epsilon_{n}-\epsilon_{n'} -
\hbar\omega) \label{Sab}
 \end{eqnarray}

\begin{figure}
     \begin{center}
         \includegraphics[height=3.55in,width=4.75in,angle=0]{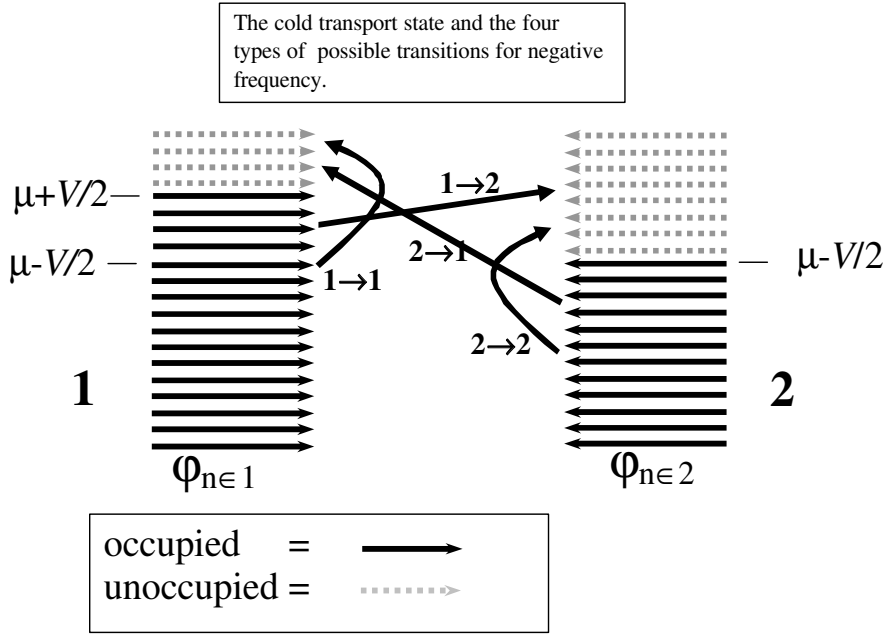}
         \caption{The Landauer model, states and transitions.}
     \end{center}
 \end{figure}

Each of these four sums can be evaluated separately by calculating
the current matrix elements in eq.(\ref{Svanhove-auto}),
transforming the sums over $k$ and $k'$ (which are implicit in the
sums over $n$ and $n'$) into integrals, performing the summation
over $i$ according to Eqs. (\ref{defi}) and (\ref{defPi}), and
integrating using the condition $\hbar\omega,eV, k_BT \ll \mu$ and
the unitarity of the scattering matrix, $\sum_{\gamma} s_{\alpha
\gamma} s_{\beta \gamma}^\ast=\delta_{\alpha \beta}$. The final
results are:
\begin{eqnarray}
 S_{1\rightarrow 2}(\omega)=
 \frac{e^2\widetilde{T}(1-\widetilde{T})}{h}\frac{\hbar\omega-eV}{e^{\beta(\hbar\omega-eV)}-1}
 \label{S12}
\end{eqnarray}
\begin{eqnarray}
 S_{2\rightarrow 1}(\omega)=
 \frac{e^2\widetilde{T}(1-\widetilde{T})}{h}\frac{\hbar\omega+eV}{e^{\beta(\hbar\omega+eV)}-1}
\label{S21}
\end{eqnarray}
\begin{eqnarray}
 S_{1\rightarrow 1}(\omega)=S_{2\rightarrow 2}(\omega)=
 \frac{e^2 \widetilde{T}^2}{h}\frac{\hbar\omega}{e^{\beta\hbar\omega}-1}.
\label{S22}
\end{eqnarray}
 Where, $\widetilde{T}\equiv |s_{21}|^2$ is the transmission from
arm 1 to 2. Substitution in eq.(\ref{partialsums}) yields:
\begin{eqnarray}
 S(\omega)=\frac{e^2\widetilde{T}(1-\widetilde{T})}{h}
 [F(\hbar\omega-eV)+F(\hbar\omega+eV)]+\frac{2e^2
 \widetilde{T}^2}{h}F(\hbar\omega)
\label{Sfinal}
 \end{eqnarray}
where  $F(x)\equiv x(e^{\beta x}-1)^{-1}$. Eq. (\ref{Sfinal}) is
the non-symmetrized power spectrum, at finite frequency (positive
or negative), voltage and temperature. It has been previously
obtained by Aguado and Kouwenhoven \cite{Leo} (they use an
opposite convention for the sign of $\omega$). Eq. (\ref{Sfinal})
is consistent with the zero-temperature limit, i.e., the cold
transport case, that was obtained by Lesovik and Loosen
\cite{LES1}. Unlike the symmetrized version that was derived by
Yang \cite{Yang} and  de Jong and Beenakker \cite{dJ-B}, here
$S(\omega)\not=S(-\omega)$.

Since the system is not in equilibrium, the detailed-balance
condition eq.(\ref{db}) is not satisfied, however, a modified
version of it does exists. To obtain it, note that
Eqs.(\ref{defPi}) and (\ref{Sab}), or (\ref{S12}),(\ref{S21}) and
(\ref{S22}) imply:

\begin{eqnarray}
 S_{1\rightarrow 1}(\omega)=
 S_{1\rightarrow 1}(-\omega)e^{-\beta\hbar\omega},  ~~~S_{2\rightarrow 2}(\omega)=S_{2\rightarrow 2}(-\omega)e^{-\beta\hbar\omega}
\end{eqnarray}
\begin{eqnarray}
 S_{1\rightarrow 2}(\omega)=
 S_{2\rightarrow 1}(-\omega)e^{-\beta\hbar\omega+\beta eV}
 ,~~~S_{2\rightarrow 1}(\omega)=
 S_{1\rightarrow 2}(-\omega)e^{-\beta\hbar\omega-\beta eV}
\  \nonumber \label{DBneq}
\end{eqnarray}
In equilibrium, $eV=0$ and the last two relations have the form of
first two and then, by eq.(\ref{partialsums}), the ordinary
detailed-balance relation, eq. (\ref{db}) is recovered. The finite
voltage creates a nonequilibrium but stationary state which is
maintained by virtue of more complicated detailed-balance
relations between upward and downward transitions, given by the
above four equations. In particular these relations imply that
$S(-\omega)\not= 0$ even at zero temperature which is not
surprising since in the cold transport state, eq.
(\ref{deftransportgas}), downward transitions are possible from
occupied states with $n\in 1$
 within the energy window $[\mu-eV/2, \mu+eV/2]$ into the empty states with $n\in 2$
 and within the same energy window. That is, emission is possible.

 Another conclusion arising from the above four equations is that in order
 to recover the classical symmetry $S(\omega)=S(-\omega)$, the condition $\hbar|\omega|\ll k_BT$
 is not longer enough and should be replaced by: $eV+\hbar|\omega|\ll k_BT$.

\section{Kubo Formula for Nonequilibrium Systems}

Here we would like to make another application of the point of
view of section \ref{general}. We again allow \cite{Yang,Kubo1}
for the antenna a large class of nonequilibrium situations - all
those in which the density matrix is diagonal in the eigenstate
basis (and it therefore commutes with the Hamiltonian). This may
include electron-electron interactions and the possibility that
$\la\hat j(t)\ra \ne 0$.

A useful example to have in mind is a Landauer-type transport
system of section 3, which has a two-terminal linear conductance,
per spin channel, of
\begin{equation}
G = \frac{e^{2}}{2\pi\hbar  }  \widetilde{T}.
\label{Land}
\end{equation}

We use the explicit exact expression of eq.\ref{Sq} for the
current power-spectrum $S(\omega)$. As emphasized in section
\ref{general}, its important physical significance is that it
gives \cite{Baym,power} the  emission rate  into the vacuum (i.e. the
state of the EM field where all $N_{\omega} = 0$) for $\omega
> 0$ and  the absorption rate for $\omega < 0$ and an EM field
with one photon, $N_{|\omega|} =  1$.

We first take the (possibly D.C. driven) "antenna" to be in a
given (usually the lowest) state consistent with the external
driving (this will be valid, for example, when the reservoirs that
feed the antenna are at $T = 0$ and there is either no coupling
with any additional thermal bath, or when that coupling is with a
bath at $T \rightarrow 0$, and no appreciable heating by the D.C.
current has taken place. In the latter case, some tendency toward
equilibrating the  chemical potentials of the left- and right-
moving electrons may occur, with a reduction of the D.C. current,
which is of no particular concern to us). In the case of the
Landauer-type system of section 3, this lowest state is the cold
transport one, defined in eq. \ref{deftransportgas}. The energy
absorption rate, $R_a (\omega )$, by the antenna from a
\textit{classical field} (with $N_\omega \gg 1 $ photons and a
negligible spontaneous component of the emitted radiation) with a
frequency $\omega
> 0$ is given via the usual treatment \cite{Baym} by:
\begin{equation}
R_a (\omega ) = \omega |A(\omega)|^2 S(-\omega) / (\hbar c^2).
\label{absrate}
\end{equation}
The emission rate, $R_e (\omega )$   is given by
\begin{equation}
R_e (\omega ) = \omega |A(\omega)|^2 S(\omega) / (\hbar c^2).
\label{emrate}
\end{equation}

The {\em net} absorption rate, $R_{a,net} (\omega )$ is given by
the difference between eqs. (\ref{absrate}) and (\ref{emrate}). It
is also given, writing  the infinitesimal "tickling" electric
field as $E(\omega) = iA(\omega)\omega/c $, by
\begin{eqnarray}
\label{abs}
R_{a,net} (\omega ) =
-2G_d(\omega) (\omega/c)^2 |A(\omega)|^2,
\end{eqnarray}
where the volume of the  system is defined as unity.

$G_d(\omega)$ is the differential ac conductance. It is defined as
the in-phase (dissipative) linear response (A.C. current) to the
tickling A.C. field at frequency $\omega$. A finite D.C. current
which in turn flows in response to the finite applied D.C. voltage
\cite{dc} $V$, is allowed.

Using eqs. \ref{absrate}, \ref{emrate} and \ref{abs}, we reach our
principal conclusion that {\em the antisymmetric part of the
current noise in our nonequilibrium system (i.e. typically
including quantum shot-noise \cite{KhlusLesovikButtiker}) is
related  to the differential ac conductance
$G_d(\omega)$ at the same frequency $\omega$.}

\begin{equation}
S(-\omega) -  S(\omega)= 2  \hbar\omega G_d(\omega),
~~~~~~~~~~~~~(\omega \ge 0), \label{fdt}
\end{equation}
Eq. \ref{fdt} is the simple but nontrivial generalization of the
Kubo  formula \cite{Kubo} for the current-carrying,
nonequilibrium, case. The antisymmetric combination appearing on
the LHS corresponds to the Fourier transform of the commutator of
the Heisenberg current operators at different times. A similar
expression was obtained by Lesovik and Loosen \cite{LES1} and by
Lesovik \cite{Usp}. Here we have interpreted it physically and put
it in a general context

It is straightforward to generalize the above treatment to the
case where the antenna is not at its lowest possible state, but
has a density matrix (assumed to be diagonal in the eigenstates'
basis) which allows the population of a number of states.
The general form of the result of  the {\em net} absorption rate
is still valid. {\em Therefore, our principal result, eq.
\ref{fdt}, is unchanged.}

We now verify the above results with the example of the
Landauer-type model, of section 3. From eq. (\ref{Sfinal}) and
(\ref{Land}) it follows that:
\begin{equation}
S(-\omega)-S(\omega)=e^2\frac{\widetilde{T}}{\pi}\omega=2\hbar\omega G,
\label{2g}
\end{equation}
in agreement with the generalized Kubo formula, eq. \ref{fdt}. To
consider a more general case we now relax the assumption that the
scattering
 matrix is energy independent and assume instead that the scale
on which it changes is of the order of $eV$, and that $eV \gg
\hbar\omega$. We emphasize that the energy dependence of the
scattering matrix must be evaluated including the self-consistent
changes in the potential of the scatterer due to the voltage $V$.
A similar derivation to the one that led to eq. (\ref{Sfinal}) and
(\ref{2g}) now gives:
\begin{equation}
S(-\omega)-S(\omega)=e^2\frac{1}{2\pi}\omega(1 - R^2(\mu - eV/2)+
\widetilde{T}^2(\mu+eV/2))
\end{equation}
where $R= 1 - \widetilde{T}$. Approximating
$\widetilde{T}(\epsilon)=\widetilde{T}(\mu-eV/2)+\lambda
(\epsilon-\mu + eV/2)+O(V^2),$ one has:

\begin{equation}
S(-\omega)-S(\omega)=e^2\frac{\widetilde{T}}{\pi}\omega(1+\lambda
eV ),
\end{equation}
where $\widetilde{T}=\widetilde{T}(\mu - eV/2)$. This, together
with eq.\ref{fdt}, is a new prediction for the low, but finite
frequency \cite{dc} dynamic conductivity, which may be different
from the slope of the DC I-V characteristics.

The generalization of the above to the case of many transport
channels, along the lines of the usual theories of quantum
shot-noise \cite{KhlusLesovikButtiker,Yang}, is straightforward.
One uses the representation where the transmission part of the
scattering matrix is diagonal. The results are expressed in terms
of the transmission eigenvalues.

When the field probing the system has a finite number,
$N_{\omega}$, of photons, the net flow of energy, including the
spontaneous process, from the system to the field is
\cite{LES1,T0}:
\begin{equation}
R_{M}(\omega)=S(\omega)(N_{\omega}+1)
-N_{\omega} S(-\omega).
\end{equation}
Using eq.\ref{Sfinal} for the quantum shot-noise, and taking $G_d
= G$, we see that the spontaneous term is unimportant for
$N_{\omega} \gg eV(1 - \widetilde{T}) / \hbar\omega.$ In the
opposite limit the sample just emits noise into the "cold
detector".

\section*{ Acknowledgements}
This research was supported by  the German-Israeli Foundation
(GIF),  the Israel Science Foundation, and the Israeli Ministry of
Science and the French Ministry of Research and Technology.
Instructive discussions with the late R. Landauer, and with D.
Cohen, C. Glattli, M. Heiblum, J. Kurchan, R. de Picciotto and A.
Stern  are gratefully acknowledged. Some of this work was done
while UG and YI were visiting the Ecole Normale Superieure in
Paris, they thank S. Haroche and M. Voos for their hospitality and
the Chaires Internationales Blaise Pascal for support.


\begin{thebibliography}{99}
\bibitem{jq}
The relevant quantity is often the space average, because of the
excellent dipole approximation. An example \cite{Baym} is a system
whose size is much smaller than the radiation wave-length, coupled
to the EM field. In section 3, the space averaging is taken only
on the portion of the antenna which is assumed to interact with
the measuring device.
\bibitem{power}
The transition rate contains the properties of the system with
which energy is exchanged. E.g, a factor $A^2$ for the EM field
(eqs. \ref{absrate} and \ref{emrate}). Due to the factor
\cite{Baym} of $1/\omega$ in $A^2$,  $S(\omega)$  and the energy
emission rate times the photon number have the same
$\omega$-dependence.
\bibitem{dc}
The dissipated power is the time-averaged product of the driving
field and the resulting current. If one of those is D.C. and the
other has finite $\omega$, the average power vanishes.
\end{thebibliography}
\end{document}